\documentclass[twocolumn,showpacs,aps,prl,superscriptaddress]{revtex4}

\usepackage{graphicx}
\usepackage{dcolumn}
\usepackage{amsmath}
\usepackage{epsfig}

\input babarsym
\newif\ifdimspec
\def\figsize#1{\dimspecfalse \checkdim#1\end
\ifdimspec
  \def\figureWidth{#1}%
\else
  \def\figureWidth{#1 in}\fi}
\def\checkdim#1{\ifx#1\end \let\next=\relax
  \else \ifcat#1a \dimspectrue \fi \let\next=\checkdim\fi \next}
\newcommand{\lblcaption  }[2]{\caption{#2\label{\secname#1}}}
\newcommand{\twoAngFiguresEPS}[6]
{
\begin{figure}
\begin{center}
\figsize{#4}
\begin{minipage}[t]{3.2in}
\begin{center}
\epsfig{file=\sectiondir/#1.eps,width=\figureWidth,angle=#5}
\end{center}
\end{minipage}
\begin{minipage}[t]{3.2in}
\begin{center}
\epsfig{file=\sectiondir/#2.eps,width=\figureWidth,angle=#6}
\end{minipage}
\end{center}
\lblcaption{#1}{#3}
\end{figure}
}

\newcommand{\SLACPubNumber} {13556}

\def\Btag       {\ensuremath{\B_{\rm{tag}}}\xspace}
\def\bmunu      {\ensuremath{\Bp \to \mup \num}\xspace}
\def\benu       {\ensuremath{\Bp \to \ep \nue}\xspace}
\def\btaunu     {\ensuremath{\Bp \to \taup \nut}\xspace}
\def\blnu       {\ensuremath{\Bp \to \ellp \nul}\xspace}

\def\babar{\mbox{\slshape B\kern-0.1em{\smaller A}\kern-0.1em
    B\kern-0.1em{\smaller A\kern-0.2em R}}}

\def\figurebox#1#2#3{%
    \def\arg{#3}%
    \ifx\arg\empty
    {\hfill\vbox{\hsize#2\hrule\hbox to #2{\vrule\hfill\vbox to #1{\hsize#2\vfill}\vrule}\hrule}\hfill}%
    \else
    {\hfill\epsfbox{#3}\hfill}%
    \fi}

\long\def\inst#1{\par\nobreak\kern 4pt\nobreak
    {\it #1}\par\vskip 10pt plus 3pt minus 3pt}

\begin{document}

%\preprint{\babar-PUB-\BABARPubYear/\BABARPubNumber} 
%\preprint{SLAC-PUB-\SLACPubNumber} 

\begin{flushleft}
SLAC-PUB-\SLACPubNumber\\
\end{flushleft}
%\babar-PUB-\BABARPubYear/\BABARPubNumber\\
%hep-ex/\LANLNumber\\[10mm]
%\end{flushleft}
\title{
{\large \bf
Search for the Rare Leptonic Decays \boldmath{\blnu} ($\ell=e,\mu$)}
}

% Dummy author list; contact ../Pubboard Chair for current author list
%% author list as of 04-Feb-2009 (493 authors)
%
\author{B.~Aubert}
\author{Y.~Karyotakis}
\author{J.~P.~Lees}
\author{V.~Poireau}
\author{E.~Prencipe}
\author{X.~Prudent}
\author{V.~Tisserand}
\affiliation{Laboratoire d'Annecy-le-Vieux de Physique des Particules (LAPP), Universit\'e de Savoie, CNRS/IN2P3,  F-74941 Annecy-Le-Vieux, France}
\author{J.~Garra~Tico}
\author{E.~Grauges}
\affiliation{Universitat de Barcelona, Facultat de Fisica, Departament ECM, E-08028 Barcelona, Spain }
\author{M.~Martinelli$^{ab}$ }
\author{A.~Palano$^{ab}$ }
\author{M.~Pappagallo$^{ab}$ }
\affiliation{INFN Sezione di Bari$^{a}$; Dipartimento di Fisica, Universit\`a di Bari$^{b}$, I-70126 Bari, Italy }
% ...
\collaboration{The \babar\ Collaboration}
\noaffiliation

\date{\today}% It is always \today, today, but you may specify any date with \date.

\begin{abstract}
We have performed a search for the rare leptonic decays \blnu ($l=e,\mu$), using data 
collected at the \FourS resonance by the \babar\ detector at the \pep2\ storage ring.
In a sample of 468 million \BB pairs we find
no evidence for a signal and set an upper limit on the branching 
fractions $\BR(\bmunu) < 1.0\times10^{-6}$  and $\BR(\benu) < 1.9\times10^{-6}$ 
at the 90\% confidence level, using a Bayesian approach. 
\end{abstract}

\pacs{13.20.-v, 13.25.Hw}% PACS, the Physics and Astronomy Classification Scheme.

\maketitle

% text goes here

% reset footnote counter
\setcounter{footnote}{0}

%%%%%%%%%%%%%%%%%%%%%%%%%%%
%\section{Introduction}
%\label{sec:Introduction}
%%%%%%%%%%%%%%%%%%%%%%%%%%%

\begin{figure}[!htb]
\begin{center}
\includegraphics[width=4.3 cm]{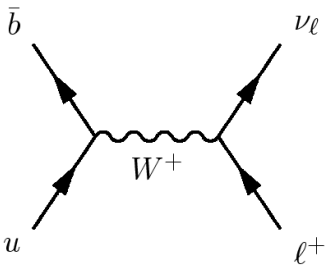}
\end{center}
\caption{\label{fig:diagram}%                                                                                                                     
Lowest order SM Feynman diagram for the purely leptonic decay $B^+ \to l^{+} \nu_{l}$.}
\end{figure}

In the standard model (SM), the purely leptonic $B$ meson decays 
\blnu~\cite{charge}
proceed at lowest order through the annihilation diagram 
shown in Fig.~\ref{fig:diagram}. 
The SM branching fraction can be  calculated as~\cite{Silverman:1988gc}

\begin{equation}
 \BR(B^+ \rightarrow \ell^+ \nu_{\ell}) = \frac{G_{F}^{2} m_{B} m_{\ell}^{2}} {8\pi} 
 \biggl( 1- \frac{m_{\ell}^{2}}{m_{\B}^{2}} \biggr)^{2} f_{B}^{2} |V_{ub}|^{2} 
 \tau_{\B},
\end{equation}

where $G_F$ is the Fermi coupling constant, $m_{\ell}$ and $m_B$ are 
respectively the lepton 
and $B$ meson masses, and $\tau_B$ is the $B^+$ lifetime. The decay rate 
is sensitive to the CKM
matrix element $|V_{ub}|$~\cite{Cabibbo:1963yz} and the $B$ decay constant $f_{\B}$ that describes 
the overlap of the quark wave functions within the meson. 

The SM estimate of the branching fraction for 
\btaunu is $(1.59 \pm 0.40)\times 10^{-4}$ assuming 
$\tau_B$ = 1.638$ \pm $0.011 ps~\cite{PDG}, $V_{ub}$ = (4.39$ \pm $0.33)$\times 10^{-3}$
determined from inclusive charmless semileptonic $B$ decays~\cite{Barberio:2006bi}  
and $f_B$ = 216$ \pm $22 MeV from lattice QCD calculation~\cite{Gray:2005ad}. 
%Due to helicity suppression, \bmunu and \benu are 
To a very good approximation, helicity is conserved in \bmunu and \benu decays, which are therefore
suppressed by factors $m_{\mu,e}^2/m_{\tau}^2$ with respect to \btaunu, leading to expected   
branching fractions of $\BR(\bmunu) = (5.6 \pm 0.4) \times 10^{-7}$ and
$\BR(\benu) = (1.3 \pm 0.4) \times 10^{-11}$. However, reconstruction
of \btaunu decays is experimentally more challenging than \bmunu or \benu due to the large missing momentum from
multiple neutrinos in the final state.

Purely leptonic $B$ decays are sensitive to physics beyond the SM,
where additional heavy virtual particles contribute to the annihilation processes.
Charged Higgs boson effects may greatly enhance or suppress the 
branching fraction in some two-Higgs-doublet models~\cite{Hou:1992sy}. Similarly, there may be
enhancements through mediation by leptoquarks in the Pati-Salam model of quark-lepton
unification~\cite{Valencia:1994cj}. 
Direct tests of Yukawa interactions in and beyond the
SM are possible in the study of these decays, as annihilation processes proceed through the
longitudinal component of the intermediate vector boson.
In particular, in a SUSY scenario at large $\tan \beta$, non-standard effects in helicity-suppressed 
charged current interactions are potentially observable, being strongly $\tan \beta$-dependent
and leading to~\cite{Hou:1992sy}:
\begin{eqnarray}
\frac{\BR(B^{+} \rightarrow l^{+} \nu_{l})_{\rm{exp}}}{\BR(B^{+} \rightarrow l^{+} \nu_{l})_{\rm{SM}} } \approx ( 1-  \tan^2 \beta \frac{m_B^2}{M_H^2} )^2 .
\end{eqnarray}

Evidence for the first purely leptonic \B decays has recently been presented by both the \babar\ and Belle collaborations.
The latest HFAG world average of the \babar\ \cite{Aubert:2007xj} and Belle~\cite{Adachi:2008ch} results
is   $\BR(\btaunu) = (1.51 \pm 0.33)\times 10^{-4}$~\cite{HFAG}.
The current best published upper limits 
on \bmunu and \benu are $\BR(\bmunu) < 1.7 \times 10^{-6}$
and $\BR(\benu) < 9.8 \times 10^{-7}$ at 90$\%$ confidence level from Belle using
a data sample of 235 \invfb~\cite{Satoyama:2006xn}.

%%%%%%%%%%%%%%%%%%%%%%%%%%%%%%%%%%%%%%%%%%%
%\section{The \babar\ detector and dataset}
%\label{sec:babar}
%%%%%%%%%%%%%%%%%%%%%%%%%%%%%%%%%%%%%%%%%%%

The analysis described in herein is based on 
the entire dataset collected
with the \babar\ detector~\cite{babar} at the \pep2 storage ring
at the \FourS resonance  (``on-resonance''), which 
consists of 468 million \BB pairs, corresponding to
an integrated luminosity
of 426 \invfb. 
In order to study background from continuum events such as $e^+ e^- \to q \bar{q}$ ($q=u,d,s,c$)
and $e^+ e^- \to \tau^+ \tau^-$, an additional sample of about 41 \invfb was collected
at a center-of-mass (c.m.) energy about 40 \mev below the \FourS resonance (``off-resonance'').

In the \babar\ detector, charged particle trajectories
 are measured with a 5-layer double-sided silicon vertex tracker and
a 40-layer drift chamber, which are contained in the 1.5 T magnetic field of a 
superconducting solenoid. 
A detector of internally reflected Cherenkov radiation provides identification of charged
kaons and pions. 
The energies and trajectories of neutral particles are measured by an electromagnetic calorimeter  
consisting of 6580 CsI(Tl) crystals. 
The flux return of the solenoid is instrumented with resistive plate chambers and, more recently, limited streamer
tubes~\cite{Benelli:2006pa}, in order to provide muon identification. 
A {\tt GEANT}4-based~\cite{geant4} Monte Carlo (MC) simulation 
of generic $B\bar{B}$, 
$q\bar{q},\,q=u,d,s,c$, and $\tau^+\tau^-$ events as well as 
$B^+\to\mu^+\nu_\mu$ and $B^+\to e^+\nu_e$ signal events is used to model
the detector response and test the analysis technique.

%%%%%%%%%%%%%%%%%%%%%%%%%%%
%\section{Analysis method}
%\label{sec:Analysis}
%%%%%%%%%%%%%%%%%%%%%%%%%%%

The \blnu decay produces a mono-energetic charged lepton in the \B rest frame 
with a momentum $p^{*} \approx m_B/2$. The \B mesons produced in \FourS decays
have a c.m. momentum of about 320 \mevc, so we initially select lepton candidates with c.m. momentum
2.4 $< p_{\rm{c.m.}} <$ 3.2 \gevc, to take into account the smearing due to 
the motion of the $B$. A tight particle identification requirement is applied to the candidate
lepton in order to discard fake muons or electrons.

Since the neutrino produced in the signal decay is not detected, 
all charged tracks besides the signal lepton and all neutral energy
deposits in the calorimeter are combined to reconstruct the 
companion (tag) \B.
We include all neutral calorimeter clusters with 
cluster energy greater than 30 \mev. Particle identification is applied to the charged tracks 
to identify electrons, muons, pions, kaons and protons in order to assign the most likely mass 
hypothesis to each \Btag daughter and thus improve the reconstruction of the \Btag.
Events which have additional lepton candidates 
are discarded. These typically arise from semileptonic \Btag or 
charm decays and indicate the presence of additional neutrinos, for which
the inclusive \Btag reconstruction is not expected to work well. 

The signal lepton's momentum in the signal $B$ rest frame $p^\ast$ is refined 
using the \Btag momentum direction. We assume that
the signal \B has a c.m. momentum
of 320 \mevc and choose its direction as opposite that of the
reconstructed \Btag to boost the lepton candidate into the signal \B rest frame.

Signal events are selected using the kinematic variables \DeltaE $ = E_{B}-E_{\rm beam}$ ,
where $E_B$ is the energy of the \Btag and $E_{\rm beam}$ is the beam energy, all in the c.m. frame. 
For signal events in
which all decay products of the \Btag are reconstructed, we expect the $\Delta E$ distribution 
to peak near zero. However, we are often unable to reconstruct all \Btag decay
products, which biases the $\Delta E$ distribution toward negative values.
For continuum backgrounds, $\Delta E$ is shifted toward relatively large positive
values since too much energy is attributed to the nominal \Btag decay, while there is
a negative bias in $\tau^+\tau^-$ events due to the unreconstructed neutrinos.

We require the tag $B$ to satisfy $-$2.25 $< \Delta E <$ 0 GeV for
\bmunu decays. For \benu decays, we require a linear combination of $\Delta E$ and the tag \B
transverse momentum $p_{T}$ to satisfy $(p_{T} + 0.529 \cdot \Delta E)<$0.2 and $(p_{T} - 0.529 \cdot \Delta E)<$1.5. 
This selection rejects background events arising from 
two-photon process $e^+e^-\to
e^+e^-\gamma^\ast\gamma^\ast,\; \gamma^\ast\gamma^\ast\to hadrons$ with
one of the final electrons scattered at a large angle and detected.
The coefficient of the $\Delta E$ term is extracted from the data.

Backgrounds may arise from any process producing charged tracks in the 
momentum range of the signal, particularly if the charged tracks are  
leptons. The two most significant backgrounds are \B semileptonic decays involving
$b\rightarrow u l \nu_{l}$ transitions in which the 
momentum of the leptons at the endpoint of the spectrum approaches that of the signal,
and from continuum and $\tau^+ \tau^-$ events in which a charged pion 
is mistakenly identified as a muon or an electron.

\begin{figure}[t!]
\begin{center}
\includegraphics[width=4.3cm]{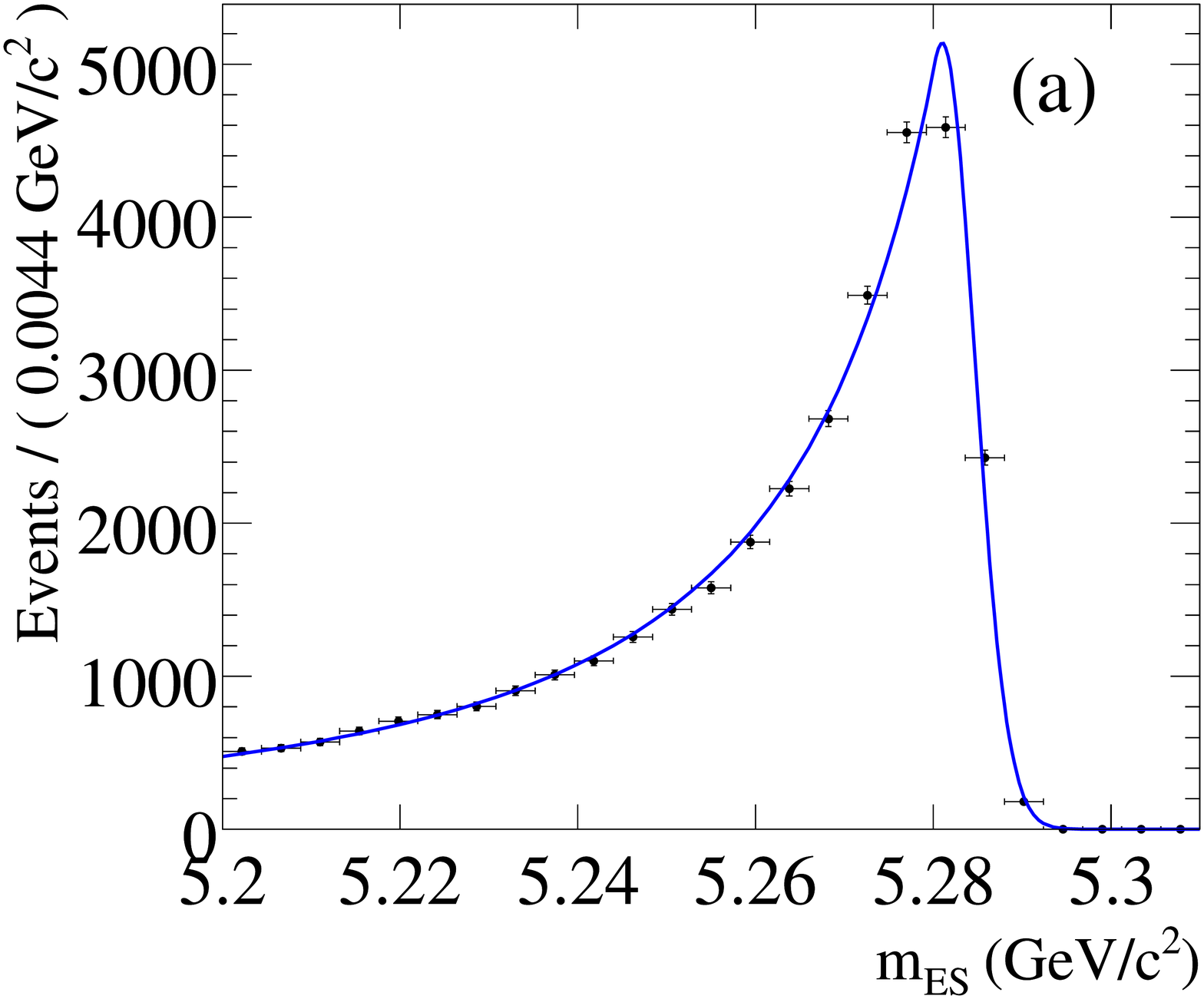}
\hspace{-0.3cm}
\includegraphics[width=4.3cm]{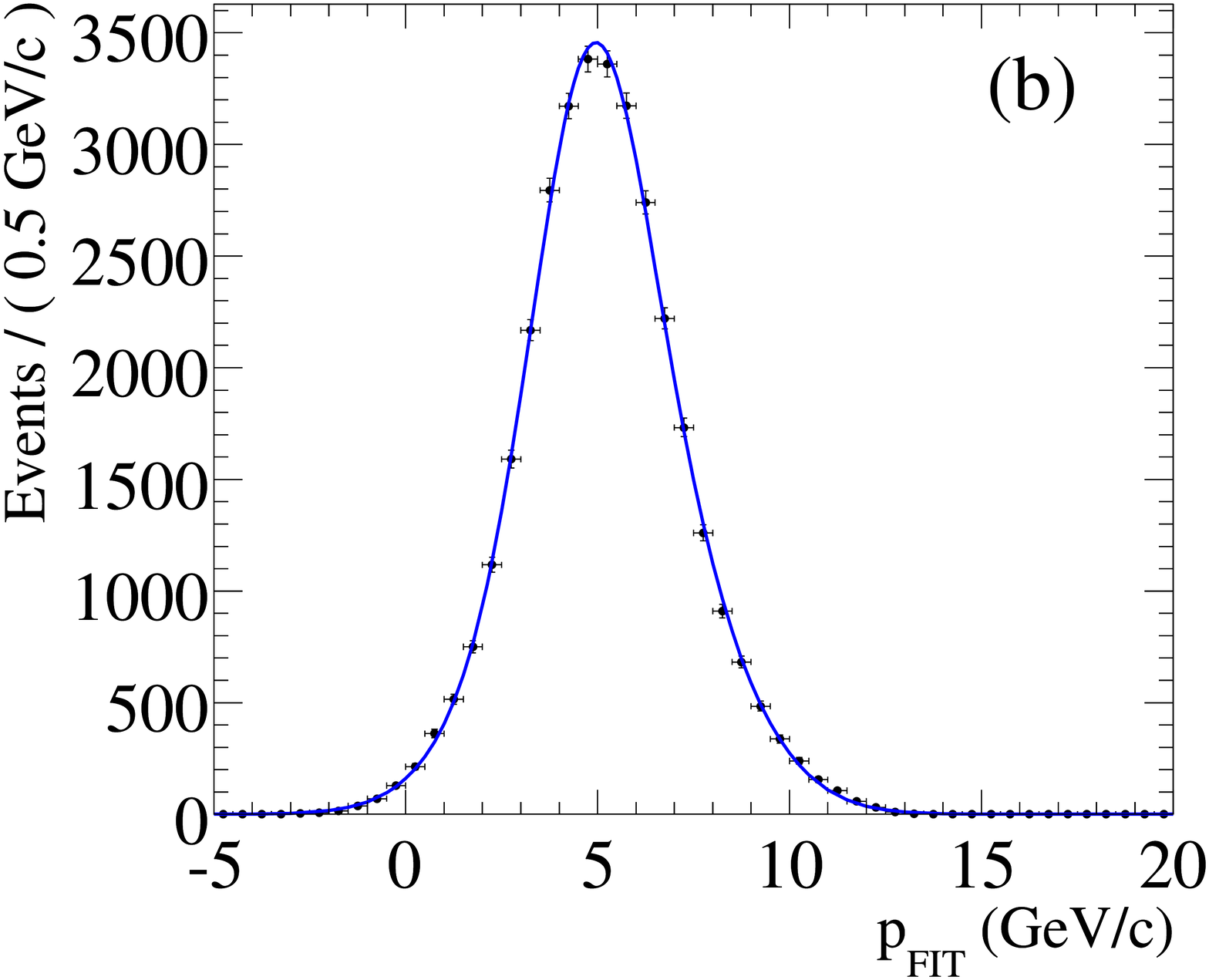}\\
\hspace{-0.3cm}
\includegraphics[width=4.3cm]{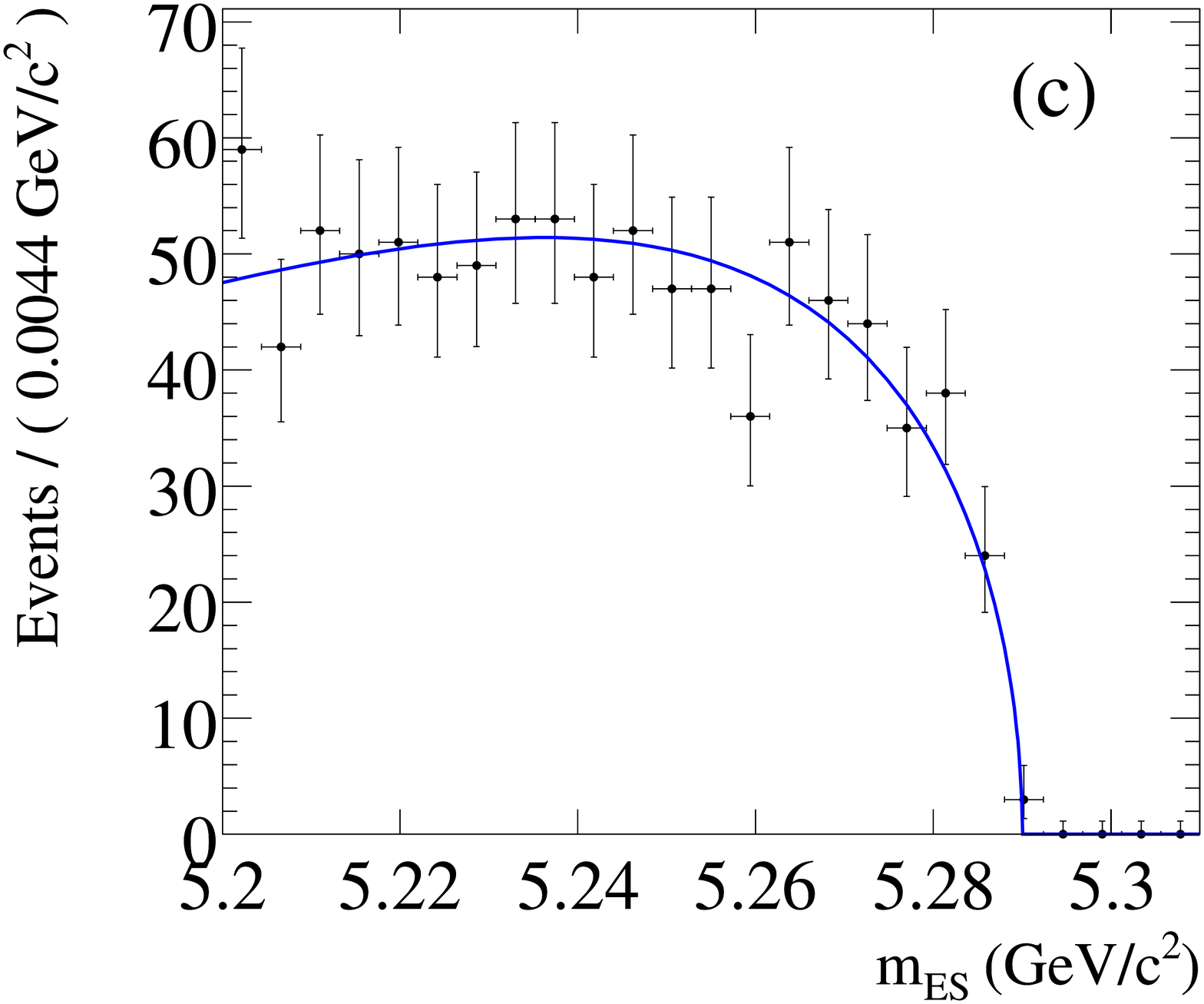}
\hspace{-0.3cm}
\includegraphics[width=4.3cm]{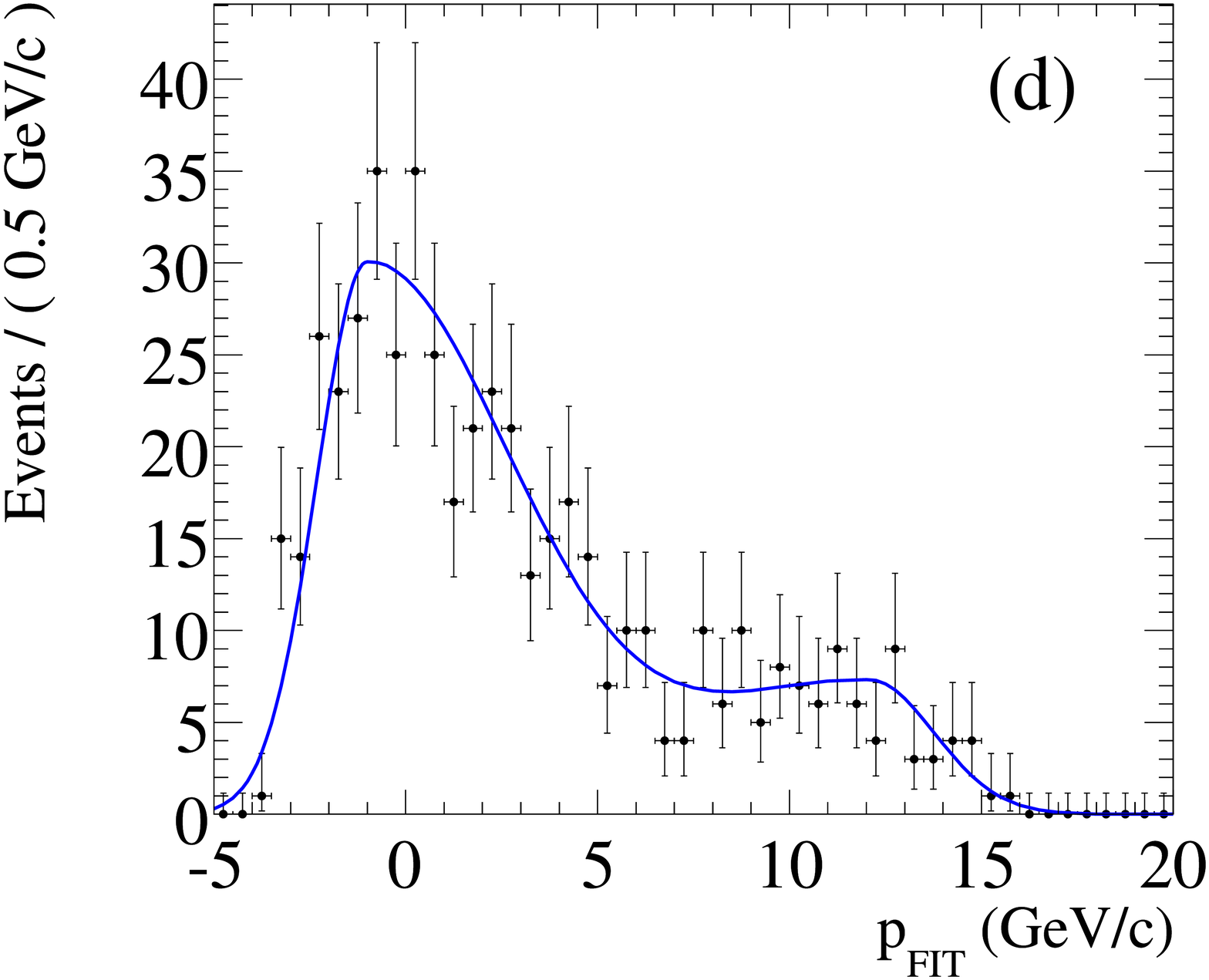}\\
\caption{Distributions of signal (a,b) and background (c,d) $m_{ES}$ (left) and $p_{\rm{FIT}}$ (right)  for \bmunu from MC simulation (a,b and c) and from 
$m_{ES}$ sideband 5.17  $<m_{ES}<$  5.2 GeV/c$^2$ (d).}
\label{fig:parmu}
\end{center}
\end{figure}

\begin{figure}[thb!]
\begin{center}
\includegraphics[width=4.3cm]{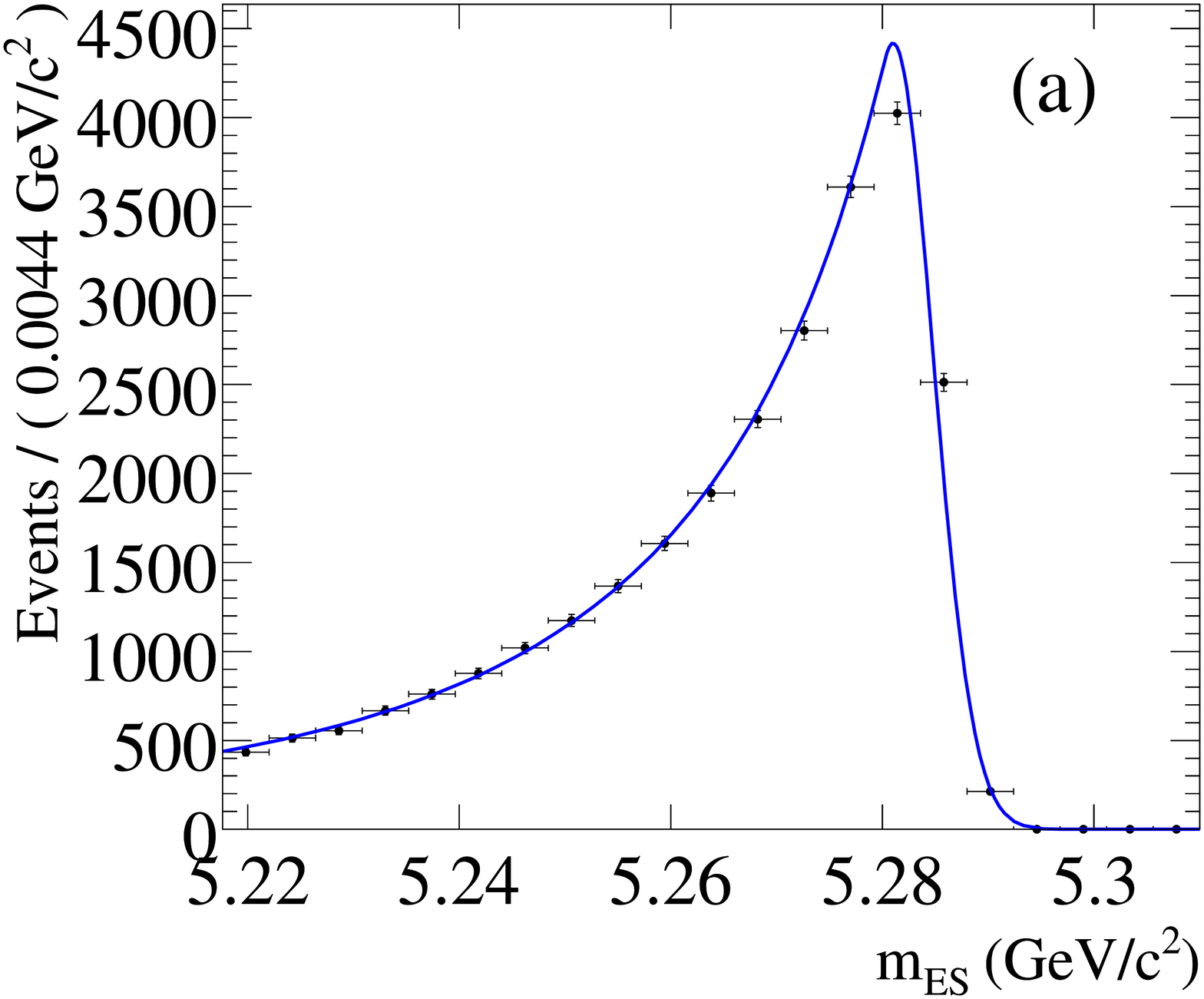}
\hspace{-0.3cm}
\includegraphics[width=4.3cm]{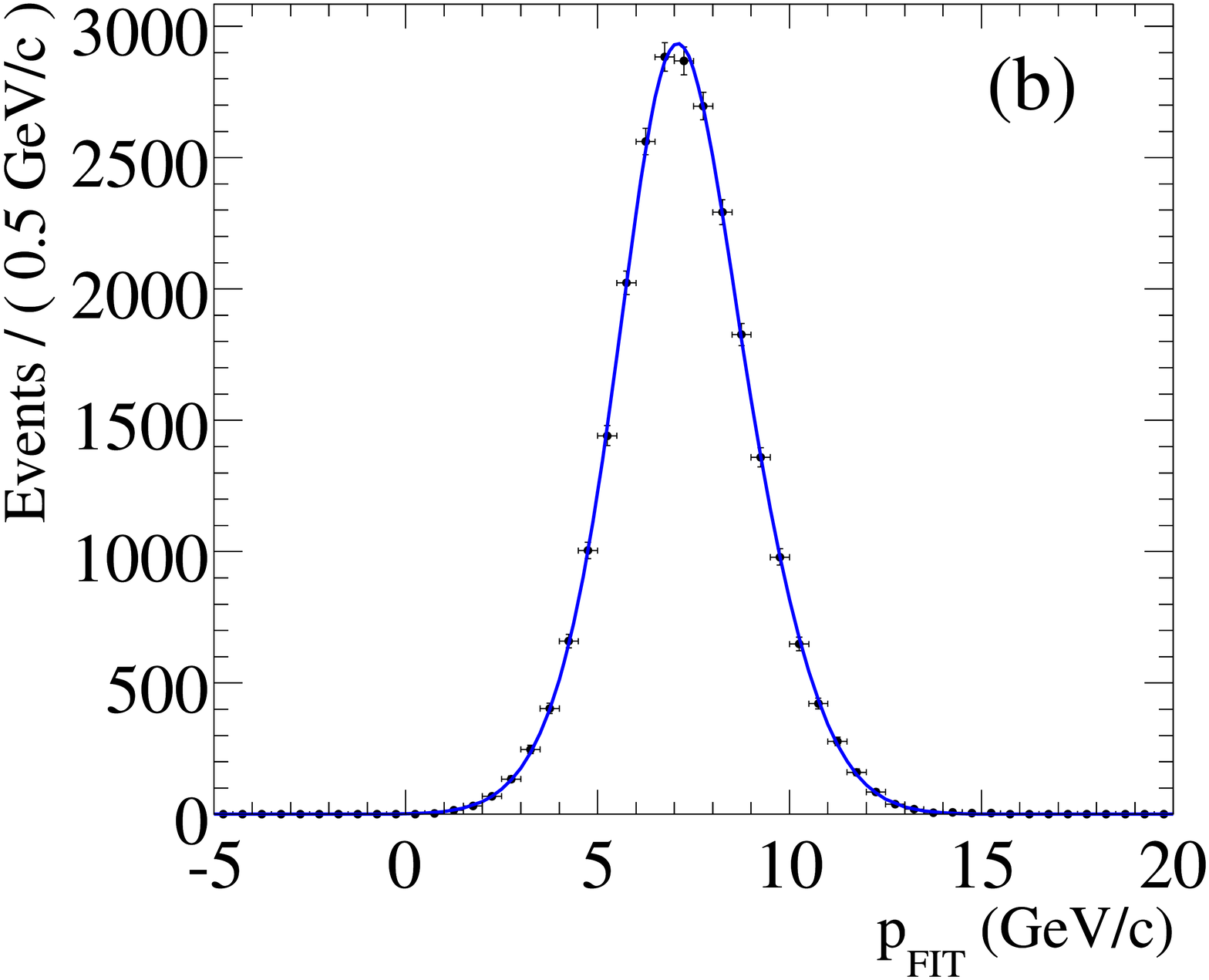}\\
\hspace{-0.3cm}
\includegraphics[width=4.3cm]{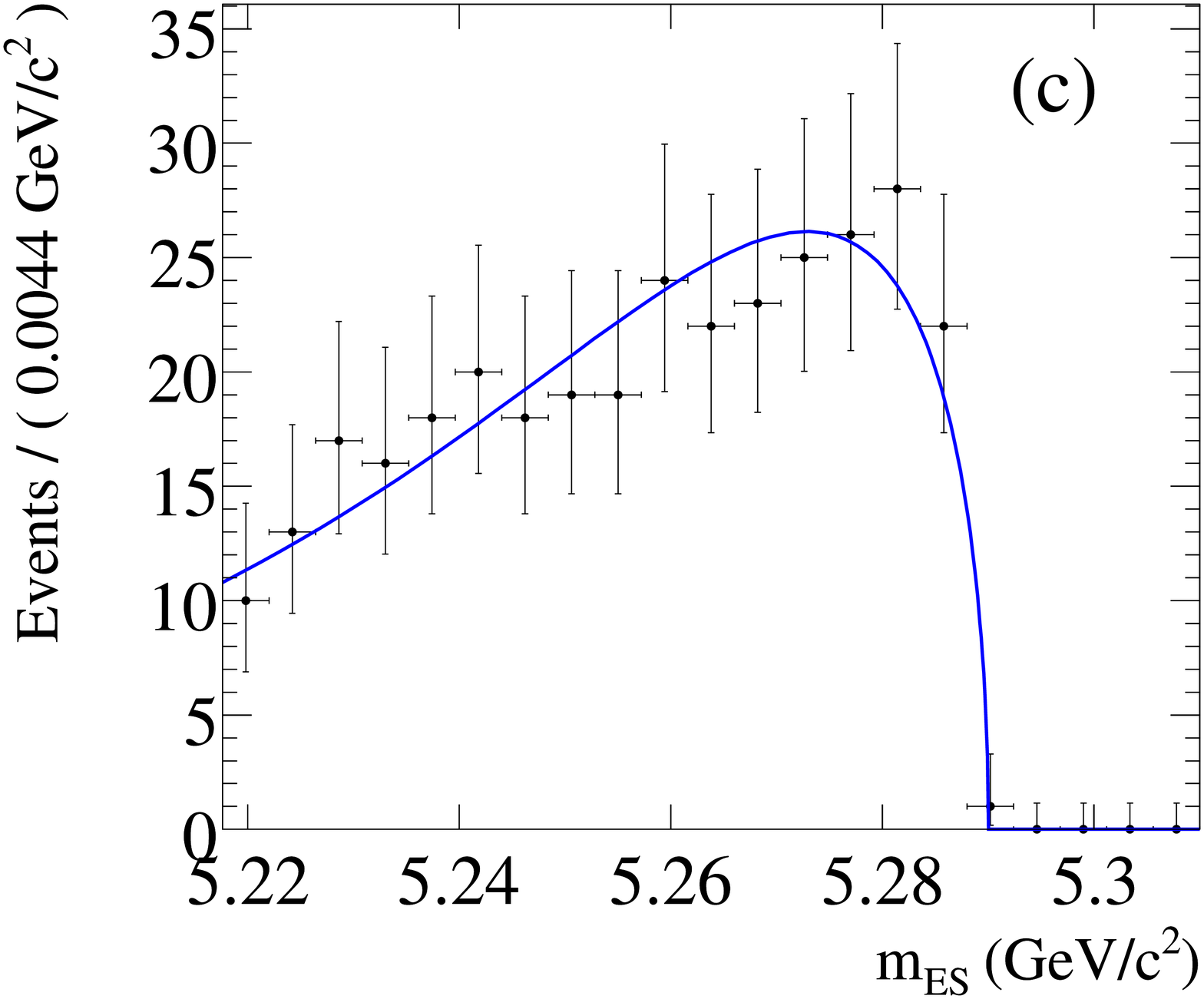}
\hspace{-0.3cm}
\includegraphics[width=4.3cm]{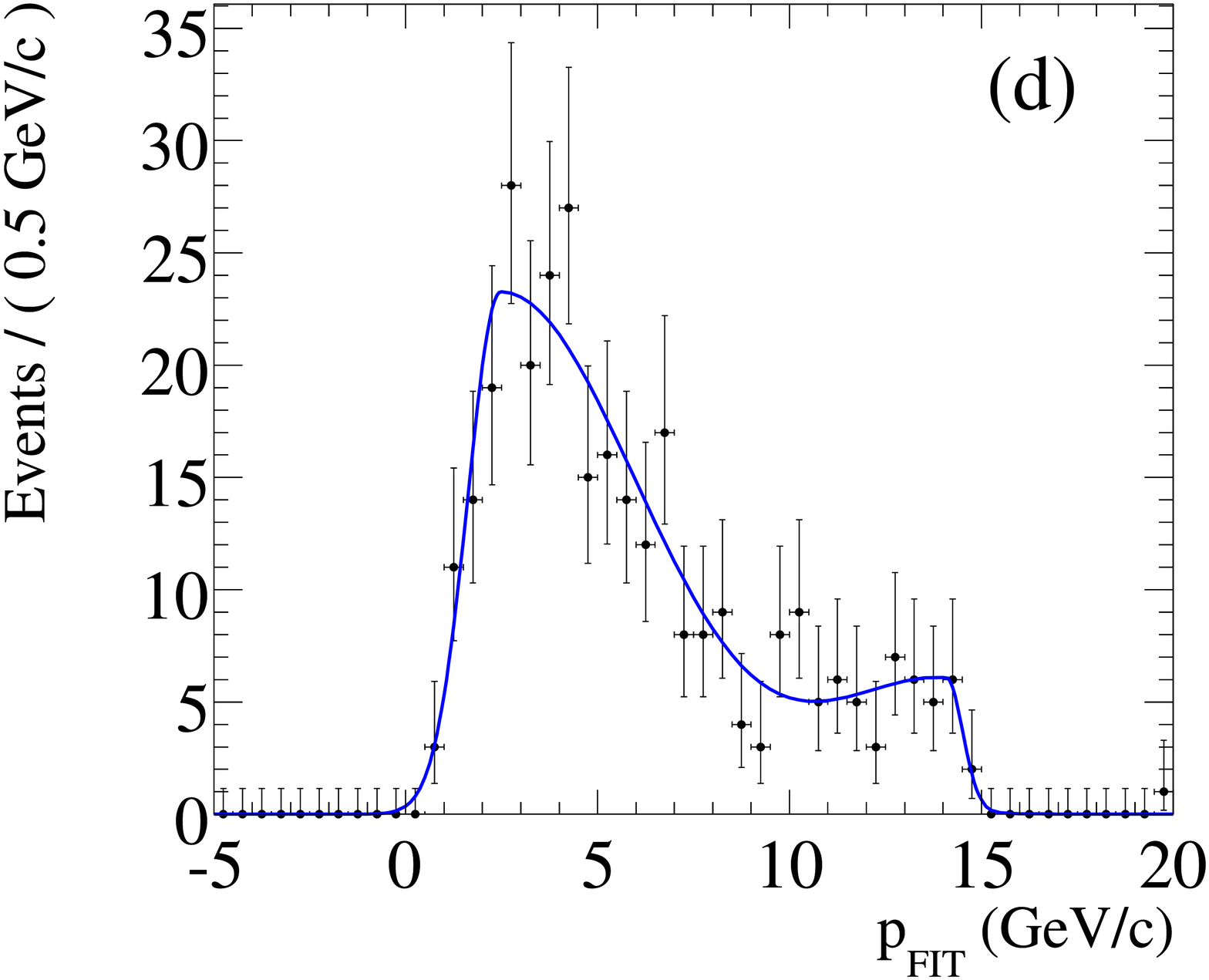}\\
\caption{Distributions of signal (a,b) and background (c,d) $m_{ES}$ (left) and $p_{\rm{FIT}}$ (right)  for \benu from MC simulation.}
\label{fig:parel}
\end{center}
\end{figure}

Continuum events tend to produce a jet-like event topology,
while \BB events tend to be more isotropically distributed
in the c.m. frame, and are suppressed using event shape parameters.
Five different spatial and kinematical variables, considered 
separately for \bmunu and \benu, 
are combined in Fisher discriminants~\cite{fisher}.
The most effective discriminating parameters are the ratio of the second $L_2$ and the zeroth $L_0$
monomial $ L_n = {\large \Sigma}_i |{\vec p}_i| \cos(\alpha)^n$,
where the sum runs over all \Btag daughters having 
momenta ${\vec p}_i$ and 
$\alpha$ is the angle with respect to the lepton candidate momentum, both in the c.m. frame, 
and the sphericity $S = \frac{3}{2}{\rm min} \frac{\Sigma_j (p_{jT})^2}{\Sigma_j (p_j)^2 }$,
where the $T$ subscript denotes the momentum component transverse to the sphericity axis,
which is the axis that minimizes $S$.
$S$, in fact,  tends to be closer to 1 for spherical events and 0 for jet-like events.
In order to take into account the changes in detector performance
throughout the years,  in particular in muon identification, the data sample is divided into 
six different data taking periods, and the Fisher discriminants and selection
criteria are optimized separately with the algorithm described in~\cite{narsky}
for each period.%~\cite{epaps}. 

The two-body kinematics of the signal decay is exploited by combining the signal lepton
momentum in the \B rest frame $p^*$ and $p_{\rm{c.m.}}$
in a second Fisher discriminant ($p_{\rm{FIT}}$) which discriminates against the remaining
semileptonic $b \bar{b}$ and continuum background events which populate the end of
the lepton spectrum in both frames. The $p^*$ and $p_{\rm{c.m.}}$ coefficients in the linear
combination are determined separately for \bmunu and \benu with~\cite{narsky}.

We employ an extended maximum likelihood (ML) fit to extract signal and background yields using
simultaneously the distributions of the 
Fisher output $p_{\rm{FIT}}$ and the energy-substituted mass \mes, defined as
$\sqrt{E_{\rm beam}^{2}-|\vec{p}_B|^{\;2}}$, where $\vec{p}_B$ is the momentum of the
reconstructed \Btag candidate in the c.m. frame.

Signal $m_{ES}$ and $p_{\rm{FIT}}$ probability density functions (PDFs)
are fixed in the final fit and are parameterized from  simulated 
events, respectively, with a Crystal Ball function~\cite{CB} and the sum of two Gaussians 
(double Gaussian) for both \bmunu and \benu.

The background $m_{ES}$ distribution is described by an
ARGUS function whose slope is determined
in the fit to the yields~\cite{argus}.
To parameterize the background $p_{\rm{FIT}}$ distributions, we studied the possibility
of using the $m_{ES}$ sideband of on-resonance data. We found the \bmunu
sideband suited for this purpose, while the \benu sideband is not
sufficiently populated.
We use the region 5.17  $<m_{ES}<$  5.2 GeV/c$^2$ 
to parameterize the \bmunu  background $p_{\rm{FIT}}$ distribution, 
and simulated events for the background \benu  $p_{\rm{FIT}}$ distribution.

Separately for \bmunu and \benu, the sum of two
Gaussians with different sigmas on the right and the left of the mean 
(bifurcated Gaussians) is used to parameterize the background $p_{\rm{FIT}}$ distribution and the
relative fraction of the two bifurcated Gaussians is determined from the fit to the data.
Figures~\ref{fig:parmu} and \ref{fig:parel} show background and signal  $m_{ES}$ and $p_{\rm{FIT}}$ distributions
for \bmunu and \benu, respectively, with the PDFs described above superimposed.
   
%%%%%%%%%%%%%%%%%%%%%%%%%%%
%\section{Results}
%\label{sec:Results}
%%%%%%%%%%%%%%%%%%%%%%%%%%%

\begin{figure}[t!]
\begin{center}
\includegraphics[width=4.3cm]{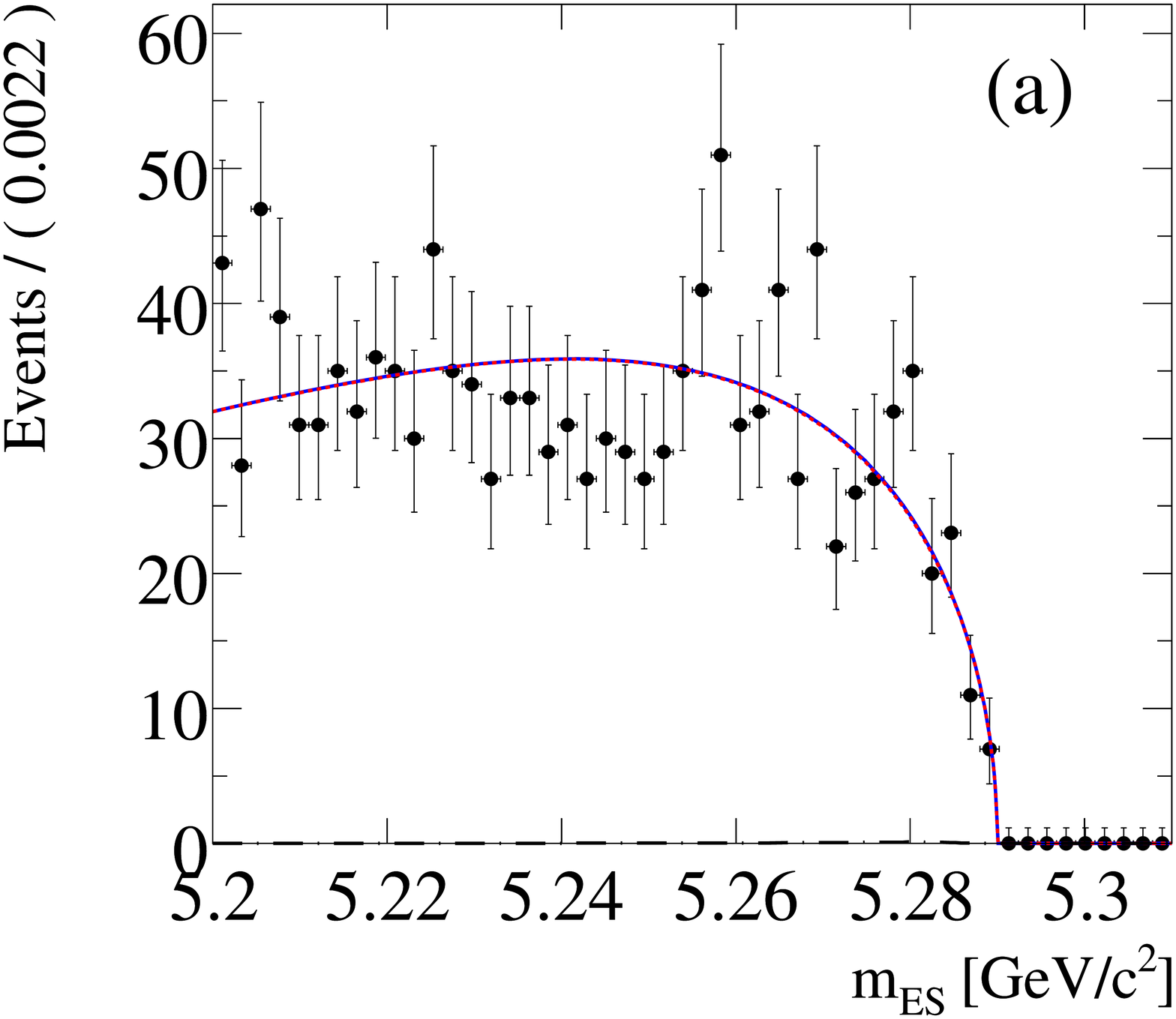}
\hspace{-0.3cm}
\includegraphics[width=4.3cm]{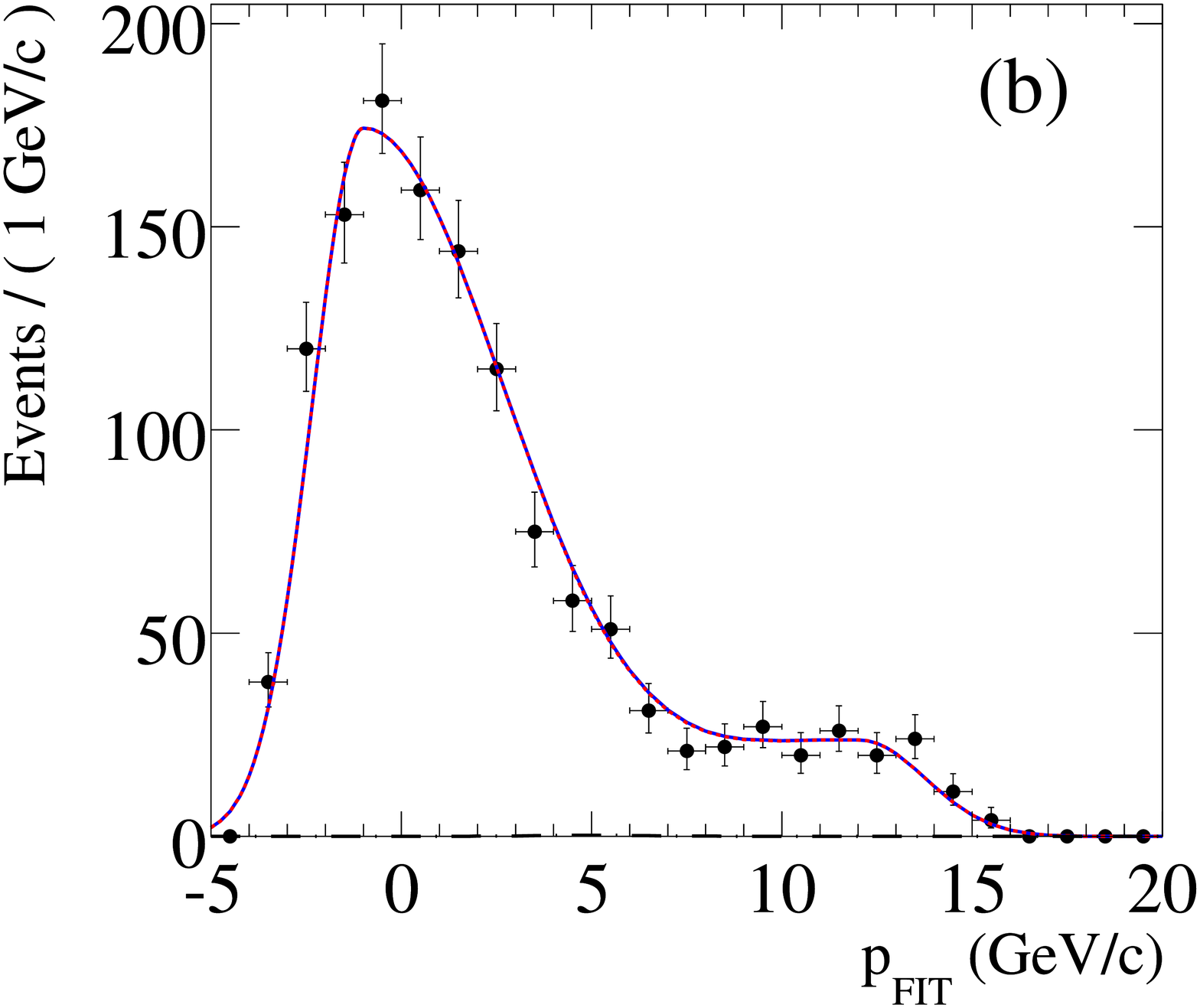}\\
\hspace{-0.3cm}
\includegraphics[width=4.3cm]{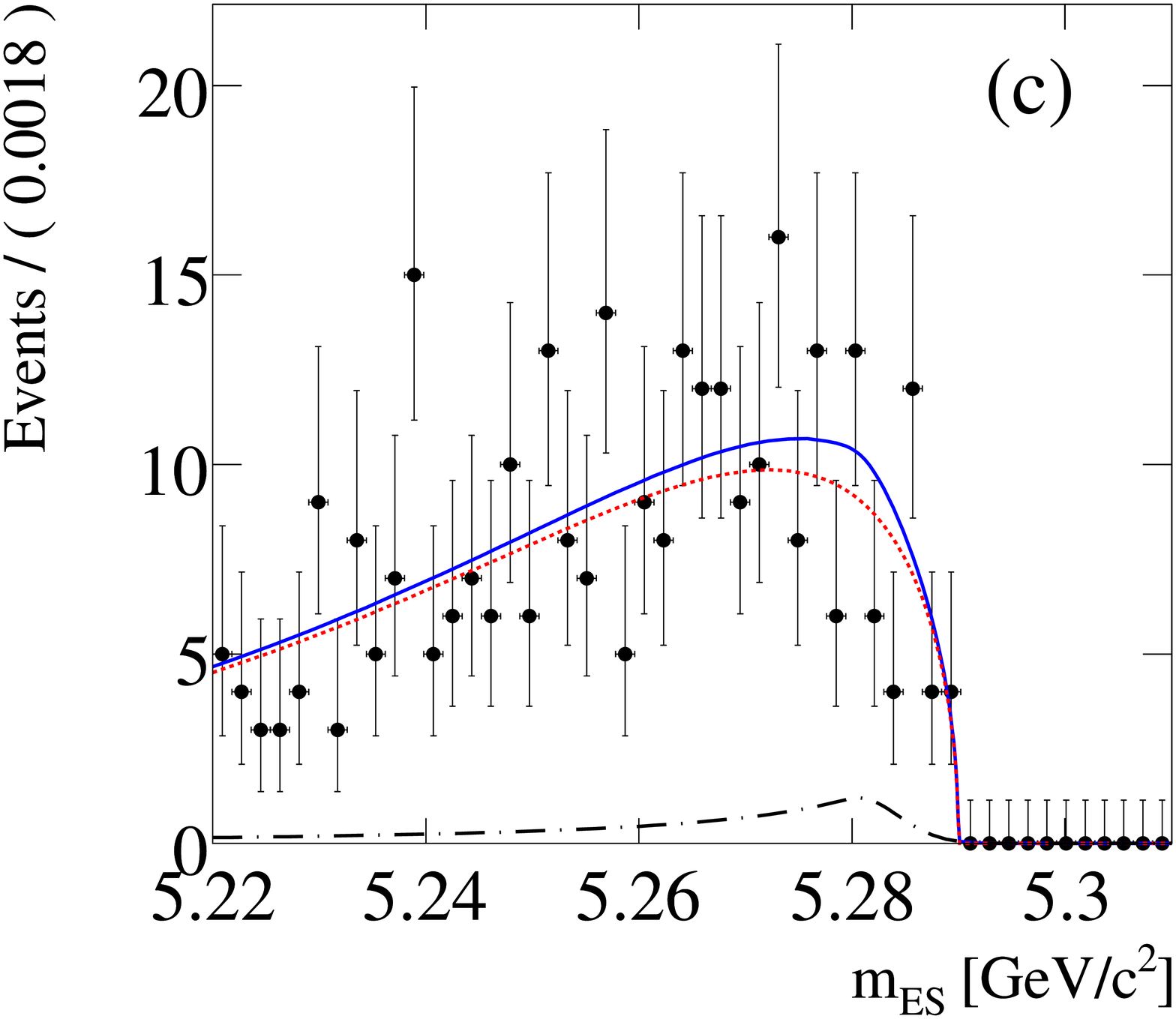}
\hspace{-0.3cm}
\includegraphics[width=4.3cm]{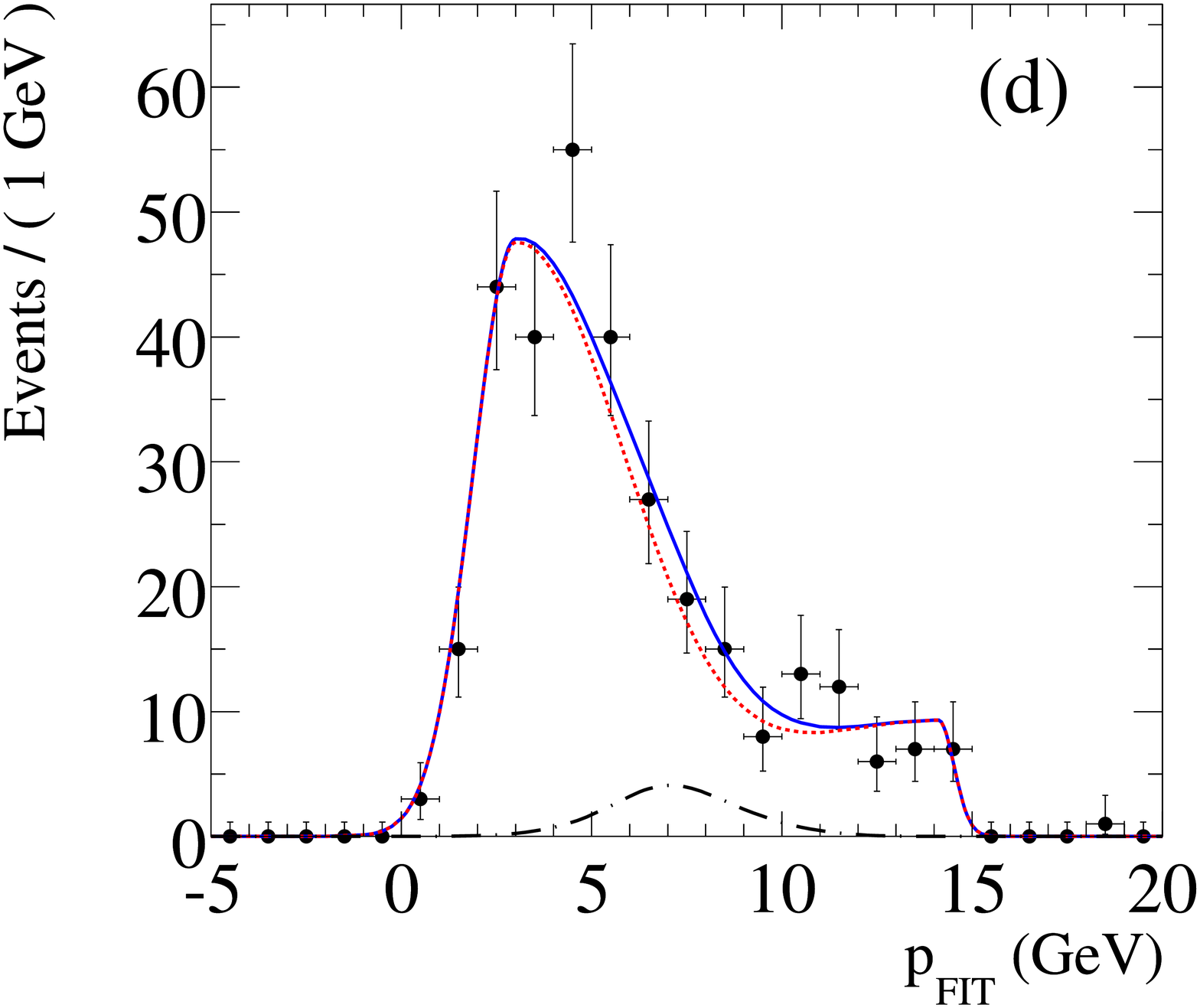}
\hspace{-0.3cm}
\caption{Final fit to the data projected on $m_{ES}$ (left) and $p_{\rm{FIT}}$ (right) distributions for \bmunu events (a,b) and \benu events (c,d)
: the solid blue line is the total PDF, the dashed red line is the background PDF and the dashed-dotted black line is the signal PDF}
 \label{fig:unblind} 
\end{center}
\end{figure}

In the on-resonance data the ML fit returns 1 $\pm$ 15 signal \bmunu candidate events and
18 $\pm$ 14 signal \benu candidate events. 
Distributions of the fit data events with the final fit
superimposed, as well as the signal and background PDFs, are shown in
Figure~\ref{fig:unblind}
for \bmunu and \benu, respectively, projected on $m_{ES}$ and $p_{\rm{FIT}}$.

%%%%%%%%%%%%%%%%%%%%%%%%%%%%%%%%%%
%\section{Systematic studies}
%\label{sec:Systematics}
%%%%%%%%%%%%%%%%%%%%%%%%%%%%%%%%%%
We next evaluate systematic 
uncertainties on the number of \Bpm in the sample, the signal efficiency  and the signal 
yield. 
The number of \Bpm mesons in the on-resonance data sample is 
estimated to be 468 $\times$ 10$^6$ with an uncertainty of 1.1\%~\cite{Aubert:2002hc},
assuming equal $B^+$ and $B^0$ production at the $\Upsilon(4S)$~\cite{Aubert:2005bq}.

The uncertainty in the signal efficiency includes the lepton candidate selection (particle identification, 
tracking efficiency and event selection Fisher requirement) as well as the reconstruction efficiency of the tag \B. 
The systematic uncertainty on the particle identification efficiency is evaluated using $e^+ e^- \to \mu^+ \mu^- \gamma$, 
$e^+ e^- \to e^+ e^- \mu^+ \mu^- $ and Bhabha event control samples derived
from the data, which are weighted to reproduce the kinematic distribution 
of the lepton signal candidate.
Comparing the cumulative signal efficiency obtained with and without these weights, 
a total discrepancy of 1.9$\%$ for \bmunu and 2.3$\%$ for \benu  
is found and this value is taken as the particle identification systematic uncertainty.
Tracking efficiency is studied employing $\tau$ decays, which must produce an odd
number of final state charged tracks because of charge
conservation.
Thus, one can determine an absolute efficiency because the number of events with
 a missing track can be measured.
The uncertainty associated
with the tracking efficiency and the data/MC discrepancy evaluated with this method are 
taken in quadrature for a total
tracking efficiency uncertainty of 0.4$\%$ per track.

In order to evaluate the systematic uncertainty associated with the
requirements on the Fisher discriminants, we compare 
data and MC Fisher distributions in the sidebands $\Delta E$>0 
for the \bmunu sample and $(p_{T} + 0.529 \cdot \Delta E)$> 0.2
for the \benu sample.
We fit the data/MC ratio with a linear function, with results 
consistent with a unitary ratio in the whole Fishers range.
We take the error on the intercept
as the systematic uncertainty on the Fisher discriminants, that is 1.4 $\%$
for \bmunu and 5.3$\%$ for \benu. 

The tag \B reconstruction has been studied with a control sample of 
$B^+\rightarrow D^{(*)0}\pi^+$ events, where the $D$ is reconstructed into $\bar{D}^{0} \rightarrow K^+ \pi^-$ 
and $D^{0} \rightarrow K^-  \pi^+ $,  and the $D^*$ into $D^{*0} \rightarrow D^0 \gamma$ or 
$D^{*0} \rightarrow D^0 \pi^0$.
These two-body decays are topologically very similar to our signal,
as the charged pion can be treated as the signal lepton and the $D^{(*)0}$ decays products
ignored to simulate the missing neutrino.
The tag \B reconstructed in the control sample thus simulates the tag \B reconstruction
in the nominal data sample.
We compare the efficiencies for our tag \B selection cuts in the $B^+\rightarrow D^{(*)0}\pi^+$
data and MC to quantify any data/MC disagreements that may affect the signal efficiency. We find a data/MC
discrepancy on $B^+\rightarrow D^{(*)0}\pi^+$ control sample of 3.0$\%$ for \bmunu decays and 0.4$\%$ for \benu decays,
and assign these as the signal efficiency uncertainty arising from the tag \B selection.

A summary of the systematic uncertainties in the signal efficiency is given in Table~\ref{tab:systematics_eff}. 
The final \bmunu signal efficiency is (6.1 $\pm$ 0.2)\%  and the \benu signal efficiency is (4.7 $\pm$ 0.3)\%,
where the errors are the sum in quadrature of statistical and systematic uncertainties.     
\begin{table}[!t]
 \caption{ Contributions to the systematic uncertainty on the signal efficiency. Total systematic represent the sum in quadrature of the table entries.}
 \begin{center}
 \begin{tabular}{ccc} 
\hline \hline 
   Source                         & \bmunu & \benu \\
\hline \hline
    Particle identification              & 1.9\% & 2.3 \%         \\
    Tracking efficiency                  & 0.4\% & 0.4 \%        \\
    Tag \B reconstruction                & 3.0\% & 0.4 \%        \\
    Fisher selection                     & 1.4\% & 5.3 \%        \\
\hline
   Total                                 & 3.8\% & 5.8 \%     \\
\hline 
 \end{tabular}
 \end{center}
 \label{tab:systematics_eff}
\end{table}

The systematic uncertainty in the yields comes from the $p_{\rm{FIT}}$ and $m_{ES}$ 
PDF parameters, which are kept fixed in the final fit and, in the \benu case, from
the use of MC simulation to extract the PDF shapes. 
The fit parameters extracted from MC 
are affected by an uncertainty due to MC statistics.
In order to evaluate
the systematic uncertainty associated with the parameterization, 
the final fit has been repeated 500 times for each background and signal PDF 
parameter which is kept fixed in the final fit. 
We randomly generate the PDF parameters  assuming Gaussian errors and 
taking into account all the correlations between them.
We perform a Gaussian fit to the distribution of the number of signal events for each parameter, 
take the fitted sigma  as the systematic
uncertainty and sum in quadrature. 
The total systematic uncertainty on the signal yield
from all signal and background PDF parameters is 8 events for \bmunu and
10 events for \benu.

For the \benu sample, an additional systematic uncertainty coming from possible discrepancies
in the shape of the $p_{\rm{FIT}}$ background distribution in data and simulated events
must be accounted for. The data/MC ratio of the $p_{\rm{FIT}}$ distribution in the
$m_{ES}$ sideband 5.16  $<m_{ES}<$  5.22 GeV/c$^2$ is fit with a linear function.
The background $p_{\rm{FIT}}$ distribution shape is varied according to the fitted linear function
and its associated statistical uncertainties; the total systematic contribution from this procedure is 4 events.

To evaluate the branching fraction we use the following expression:
\begin{equation}
 \BR(B\rightarrow l^+\nu)_{UL} = 
 \frac{N_{sig}}{N_{B^{\pm}}\cdot\varepsilon},
\label{eq:ul}
\end{equation}  
where $N_{sig}$ represents the observed signal yield, $N_{B^{\pm}}$ the number of $B^+ B^-$ in the sample
(where equal production of $B^+ B^-$ and $B^0 \bar{B}^0$ is assumed) and $\varepsilon$ is the  signal efficiency. 

As we did not find evidence for signal events, we employ a Bayesian approach to set upper limits on the branching fractions. 
Flat prior in the branching fractions are assumed
for positive values of the branching fractions and Gaussian likelihoods are adopted for the observed signal yield, related
to  \BR\ by Eq.(\ref{eq:ul}). The Gaussian widths are fixed to the sum in quadrature
of the statistical and systematic yield errors. The effect of systematic uncertainties 
associated with the efficiencies, modeled by Gaussian PDFs, is taken into account as well.
We extract the following 90 $\%$
confidence level upper limits on the branching fractions:
\begin{eqnarray}
\BR(B^+\rightarrow\mu^+\nu_{\mu}) &<& 1.0 \times 10^{-6}\\
\BR(B^+\rightarrow e^+\nu_{e}) &<& 1.9 \times 10^{-6}.
\end{eqnarray}
The 95\% upper limits are  $\BR(B^+\rightarrow\mu^+\nu_{\mu}) < 1.3 \times 10^{-6}$ and $\BR(B^+\rightarrow e^+\nu_{e}) < 2.2 \times 10^{-6}$ .
This result improves the previous best published limit for \bmunu branching fraction
by nearly a factor two, to a value twice the SM prediction. The \benu result is 
consistent with previous measurements.
It should be noted that the results in~\cite{Satoyama:2006xn} are obtained using a different
statistical approach to interpret the observed number of
signal events.
The results show no deviation from the SM expectations.
%%%%%%%%%%%%%%%%%%%%%%%%%%%%%%%%%%%%%%%%%%%%
% Input the ../../pubboard acknowledgements file
%%%%%%%%%%%%%%%%%%%%%%%%%%%%%%%%%%%%%%%%%%%

We are grateful for the excellent luminosity and machine conditions
provided by our \pep2\ colleagues, 
and for the substantial dedicated effort from
the computing organizations that support \babar.
The collaborating institutions wish to thank 
SLAC for its support and kind hospitality. 
This work is supported by
DOE
and NSF (USA),
NSERC (Canada),
CEA and
CNRS-IN2P3
(France),
BMBF and DFG
(Germany),
INFN (Italy),
FOM (The Netherlands),
NFR (Norway),
MES (Russia),
MEC (Spain), and
STFC (United Kingdom). 
Individuals have received support from the
Marie Curie EIF (European Union) and
the A.~P.~Sloan Foundation.


\begin{thebibliography}{99}

\bibitem{charge}Charge conjugation is implied throughout the paper.

% its Journal name, vol, page number, year
\bibitem{Silverman:1988gc}
  D.~Silverman and H.~Yao,
  %``RELATIVISTIC TREATMENT OF LIGHT QUARKS IN D AND B MESONS AND W EXCHANGE
  %WEAK DECAYS,''
  Phys.\ Rev.\  D {\bf 38}, 214 (1988).
  %%CITATION = PHRVA,D38,214;%%

\bibitem{Cabibbo:1963yz}
  N.~Cabibbo,
  %``Unitary Symmetry and Leptonic Decays,''
  Phys.\ Rev.\ Lett.\  {\bf 10}, 531 (1963);
  %%CITATION = PRLTA,10,531;%%
  M.~Kobayashi and T.~Maskawa,
  %``CP Violation In The Renormalizable Theory Of Weak Interaction,''
  Prog.\ Theor.\ Phys.\  {\bf 49}, 652 (1973).
  %%CITATION = PTPKA,49,652;%%

\bibitem{PDG} C. Amsler {\it et al.}, Physics Letters B {\bf 667}, 1 (2008). 

\bibitem{Barberio:2006bi}E.~Barberio {\it et al.}  [Heavy Flavor Averaging Group (HFAG)],
  %``Averages of b-hadron properties at the end of 2005,''
  arXiv:hep-ex/0603003.
  %%CITATION = HEP-EX/0603003;%%
\bibitem{Gray:2005ad} A.~Gray {\it et al.}  [HPQCD Collaboration],
  %``The B meson decay constant from unquenched lattice QCD,''
  Phys.\ Rev.\ Lett.\  {\bf 95}, 212001 (2005).
  %[arXiv:hep-lat/0507015].
  %%CITATION = PRLTA,95,212001;%%
\bibitem{Hou:1992sy}  W.~S.~Hou,
  %``Enhanced charged Higgs boson effects in B- $\to$ tau anti-neutrino, mu  %anti-neutrino and b $\to$ tau anti-neutrino + X,''
  Phys.\ Rev.\  D {\bf 48}, 2342 (1993).
  %%CITATION = PHRVA,D48,2342;%%
\bibitem{Valencia:1994cj} G.~Valencia and S.~Willenbrock,
  %``Quark - lepton unification and rare meson decays,''
  Phys.\ Rev.\  D {\bf 50}, 6843 (1994).
  %[arXiv:hep-ph/9409201].
  %%CITATION = PHRVA,D50,6843;%%

\bibitem{Aubert:2007xj}
  B.~Aubert {\it et al.}  [\babar\ Collaboration],
  %``A search for B+ --> tau+ nu with Hadronic B tags,''
  Phys.\ Rev.\  D {\bf 77}, 011107 (2008).
  %[arXiv:0708.2260 [hep-ex]].
  %%CITATION = PHRVA,D77,011107;%%

\bibitem{Adachi:2008ch}
  I.~Adachi  {\it et al.} [Belle Collaboration],
  %``Measurement of B- -> tau- nu_tau-bar Decay With a Semileptonic Tagging
  %Method,''
  arXiv:0809.3834 [hep-ex].
  %%CITATION = ARXIV:0809.3834;%%

\bibitem{HFAG}
  Heavy Flavor Averaging Group (HFAG), {\tt http://www.slac.stanford.edu/xorg/hfag/index.html}.

\bibitem{Satoyama:2006xn}
  N.~Satoyama {\it et al.}  [Belle Collaboration],
  %``A search for the rare leptonic decays B+ --> mu+ nu and B+ --> e+ nu,''
  Phys.\ Lett.\  B {\bf 647}, 67 (2007).  
  %[arXiv:hep-ex/0611045].
  %%CITATION = PHLTA,B647,67;%%
  

\bibitem{babar} B.~Aubert  {\em et al.} [\babar\ Collaboration], Nucl. Instrum. Methods A {\bf 479}, 1 (2002). 

\bibitem{Benelli:2006pa}
  G.~Benelli, K.~Honscheid, E.~A.~Lewis, J.~J.~Regensburger and D.~S.~Smith,
  %``The BaBar LST detector high voltage system: Design and implementation,''
  IEEE Nucl.\ Sci.\ Symp.\ Conf.\ Rec.\  {\bf 2}, 1145 (2006).
  %%CITATION = 00499,2,1145;%%


\bibitem{geant4}
S.~Agostinelli {\it et al.}  [GEANT4 Collaboration],
%``GEANT4: A simulation toolkit,''
Nucl. Instrum. Methods A {\bf 506}, 250 (2003).

\bibitem{fisher}
  R.\ A.\ Fisher, Ann. Eugenics {\bf 7}, 179 (1936); 
  G. Cowan, \emph{Statistical Data Analysis}, (Oxford University Press, 1998), p. 51.

\bibitem{narsky}
	I. Narsky,
	%StatPatternRecognition: A C++ Package for Statistical Analysis of High   Energy Physics Data
	arXiv:physics/0507143.





\bibitem{CB}
  J.~Gaiser {\it et al.},
  %``Charmonium Spectroscopy From Inclusive Psi-Prime And J/Psi
  %Radiative
  %Decays,''
  Phys.\ Rev.\  D {\bf 34}, 711 (1986).
  %%CITATION = PHRVA,D34,711;%%


\bibitem{argus}  
 H. Albrecht {\em et al.}, [ARGUS Collaboration] Phys. Lett. B {\bf 241}, 278 (1990).

\bibitem{Aubert:2002hc} B.~Aubert {\it et al.}  [BABAR Collaboration],
  %``Study of inclusive production of charmonium mesons in $B$ decay,''
  Phys.\ Rev.\  D {\bf 67}, (2003) 032002.
  %[arXiv:hep-ex/0207097].
  %%CITATION = PHRVA,D67,032002;%%

\bibitem{Aubert:2005bq}
  B.~Aubert {\it et al.}  [\babar\ Collaboration],
  %``Measurement of the branching fraction of $\Upsilon(4S) \to B^0
  %\overline{B}^0$,''
  Phys.\ Rev.\ Lett.\  {\bf 95}, 042001 (2005).
  %[arXiv:hep-ex/0504001].
  %%CITATION = PRLTA,95,042001;%%



\end{thebibliography}
\end{document}